\documentclass[aps,prl,twocolumn,showpacs,amsmath,amssymb,amsfonts,nofootinbib,long]{revtex4}
\usepackage{epsfig,latexsym,bm}

\newcommand\GeV{\mbox{GeV}}

\newcommand\kpc{\mbox{kpc}}
\newcommand\Mpc{\mbox{Mpc}}

\newcommand\G{\mbox{G}}

\newcommand\A{\mathbf{A}}

\newcommand\x{\mathbf{x}}
\newcommand\y{\mathbf{y}}

\newcommand\kk{\mathbf{k}}

\newcommand\ee{{\boldsymbol \varepsilon}}

\newcommand\mPl{m_{\rm Pl}}

\begin{document}

\title{Inflation-Produced Magnetic Fields in $R^n F^2$ and $I F^2$ models}

\author{L. Campanelli$^{1,2}$}
\email{leonardo.campanelli@ba.infn.it}
\author{P. Cea$^{1,2}$}
\email{paolo.cea@ba.infn.it}
\author{G.L. Fogli$^{1,2}$}
\email{gianluigi.fogli@ba.infn.it}
\author{L. Tedesco$^{1,2}$}
\email{luigi.tedesco@ba.infn.it}

\affiliation{$^1$Dipartimento di Fisica, Universit\`{a} di Bari, I-70126 Bari, Italy}
\affiliation{$^2$INFN - Sezione di Bari, I-70126 Bari, Italy}

\date{February, 2008}


\begin{abstract}
We re-analyze the production of seed magnetic fields during
Inflation in $(R/m^2)^n F_{\mu \nu}F^{\mu \nu}$ and $I F_{\mu
\nu}F^{\mu \nu}$ models, where $n$ is a positive integer, $R$ the
Ricci scalar, $m$ a mass parameter, and $I \propto \eta^\alpha$ a
power-law function of the conformal time $\eta$, with $\alpha$ a
positive real number. If $m$ is the electron mass, the produced
fields are uninterestingly small for all $n$. Taking $m$ as a free
parameter we find that, for $n \geq 2$, the produced magnetic
fields can be sufficiently strong in order to seed dynamo
mechanism and then to explain galactic magnetism. For $\alpha
\gtrsim 2$, there is always a window in the parameters defining
Inflation such that the generated magnetic fields are
astrophysically interesting. Moreover, if Inflation is (almost) de
Sitter and the produced fields almost scale-invariant ($\alpha
\simeq 4$), their intensity can be strong enough to directly
explain the presence of microgauss galactic magnetic fields.
\end{abstract}


\maketitle


\section{I. Introduction}

Microgauss magnetic fields are present in any type of
galaxies~\cite{Widrow}. Up to now, astronomical observations have
not been able to clarify the nature and origin of these fields, so
that ``galactic magnetism'' is still an open issue in cosmology.

If on the one hand, the presence of magnetic fields in all
galaxies could be explained by the action of astrophysical
mechanisms operating during or after large-scale structure
formation~\cite{LSS}, on the other hand, the detection of magnetic
fields in galaxies at high redshifts~\cite{z} represent a strong
hint that magnetic fields could have been generated in the very
early Universe.
Clearly, the detection of magnetic fields in the Cosmic Microwave
Background radiation would prove the primordial origin of cosmic
magnetic fields~\cite{CMB}.
In such a case, it is very probable that they have been generated
during an inflationary epoch of the Universe, since only in this
case it is possible to generate fields correlated on comoving
scale as large as galactic ones.

Indeed, many
mechanisms~\cite{CCFT,Mazzitelli,Bamba,Generation,Prokopec},
acting during Inflation, have been proposed to produce seed fields
since the seminal paper by Turner and Widrow~\cite{Turner}.

It is worth noting that, in order to explain galactic magnetism,
one needs the presence of seed magnetic fields prior to galaxy
formation of intensity generally much less than $1\mu \G$. In
fact, when a protogalaxy collapses to form a galactic disk,
magnetic fields suffer an amplification due to magnetic flux
conservation. Moreover, due to magnetohydrodynamic turbulence
effects and differential rotation of galaxy, seed fields can be
further amplified. This last mechanism, know as ``galactic
dynamo''~\cite{Dynamo}, can be very efficient and, in principle,
allow extremely week seeds to reproduce the properties of
presently-observed galactic fields.

If dynamo is operating, the present-day intensity and correlation
length required for a successful amplification are
\footnote{It is worth noting that galactic dynamo is a
still-debated and model-dependent amplification mechanism, and
that the minimum value of the seed field used in this paper, $B
\simeq 10^{-33} \G$, corresponds indeed to an extremely efficient
dynamo action. For details see, e.g., Ref.~\cite{CCFT} and
references therein.}
$B \gtrsim 10^{-33} \G$ and $\lambda \gtrsim 10 \kpc$ while,
without dynamo, a comoving seed field $B \gtrsim 10^{-14} \G$
correlated on comoving scales of order $1 \Mpc$ is needed to
explain galactic magnetism~\cite{CCFT}. Note that, following
Ref.~\cite{Widrow}, we are assuming that the redshift at which a
protogalaxy separates from the Hubble flow to then collapse is
$z_{\rm pg} \simeq 50$.

In this paper, we re-analyze the generation of seed magnetic
fields during Inflation considering non-standard terms in the
electromagnetic Lagrangian of the form $(R/m^2)^n F_{\mu
\nu}F^{\mu \nu}$ and $I F_{\mu \nu}F^{\mu \nu}$, where $n$ is a
positive integer, $R$ the Ricci scalar, $m$ a mass parameter, and
$I(\phi,R,...)$ a scalar function depending on any scalar field,
gravity, and on any other background field.

The first model was first proposed by Turner and
Widrow~\cite{Turner} for the case $n = 1$ and then extended to the
case $n \geq 1$ by Mazzitelli and Spedalieri~\cite{Mazzitelli}.
They found that, taking $m=m_e$ ($m_e$ being the electron mass),
astrophysically interesting seed fields can be generated during
Inflation for adequate values of $n$. In Section II, we will find
a different result from Ref.~\cite{Mazzitelli}. Indeed, if
$m=m_e$, the produced fields are uninterestingly small for all
$n$. Nevertheless, taking $m$ as a free parameter, the possibility
to generate magnetic fields sufficiently strong in order to seed
dynamo mechanism is not excluded.

The second model was first investigated by Bamba and
Sasaki~\cite{Bamba} who assumed de Sitter Inflation and $I$ to be
a power-law function of the conformal time $\eta$, $I \propto
\eta^\alpha$, with $\alpha$ a negative real number. In Section
III, we will extend this model to the case of positive real
$\alpha$ for both Power-Law and de Sitter Inflation.

In Section IV we draw our conclusions.


\section{II. $R^n F^2$ models}

We consider a photon $A_{\mu}$ described by the effective
Lagrangian
\begin{equation}
\label{Lagrangian} {\cal L} = - \frac14 \, F_{\mu \nu}F^{\mu \nu}
- \frac14 \, I F_{\mu \nu}F^{\mu \nu}, \;\;\;\; I =
\left(\frac{R}{m^2}\right)^{\!\!n} \!,
\end{equation}
where $F_{\mu\nu} = \partial_\mu A_\nu - \partial_\nu A_\mu$ is
the electromagnetic field strength tensor, $R$ the Ricci scalar,
$m$ a parameter having dimension of a mass, and $n$ a positive
integer. (From now on, Greek indices range from $0$ to $3$, while
Latin ones from $1$ to $3$.)

We will work in a flat Universe described by a Robertson-Walker
metric $ds^2 = a^2(d\eta^2 - d \x^2)$, where the conformal time
$\eta$ is related to the cosmic time $t$ through $d \eta = dt/a$.
We normalize the expansion parameter so that at the present time
$t_0$, $a(t_0) = 1$.

We consider the case of Power-Law Inflation described by the
equation of state $p_{\rm tot} = \gamma \rho_{\rm tot}$ with $-1 <
\gamma < -1/3$. In this case, the Hubble parameter $H^2 = (8\pi/3)
\rho_{\rm tot}/\mPl^2$ and the Ricci scalar $R = 8\pi(1-3\gamma)
\rho_{\rm tot}/\mPl^2$ evolve as $H^2 \propto R \propto \rho_{\rm
tot} \propto \eta^{s}$,
where $s = -6(1+\gamma)/(1+3\gamma)$. Here, $\rho_{\rm tot}$
($p_{\rm tot}$) is the total energy (pressure) density during
Inflation and $\mPl \simeq 1.22 \times 10^{19} \GeV$ is the Planck
mass.

We assume that during Inflation $R \gg m^2$, so that we can
neglect the Maxwell term in Lagrangian~(\ref{Lagrangian}). Since
$R$ is a decreasing function of time, it suffices to impose that
$R_1 \lesssim m^2$ or equivalently $m \lesssim m_{\rm max}$, with
\begin{equation}
\label{mmax} \frac{m_{\rm max}}{\mPl} = \sqrt{8\pi (1-3\gamma)}
\left(\frac{M}{\mPl}\right)^{\!\!2} \!,
\end{equation}
where $M^4 = \rho_{\rm tot}(\eta_1)$ is the total energy density
at the end of Inflation. Here and in the following, the subscript
``$1$'' indicates the time when Inflation ends.

The spectrum of gravitational waves generated at Inflation is
submitted to constraints coming from Cosmic Microwave Background
analysis which requires $\rho_{\rm tot}$ to be less than about
$10^{-8} \mPl^4$ on the scale of the present Hubble
radius~\cite{Turner}. This turns in a upper limit on the value of
$M$ and $s$: $M \lesssim 10^{-2} \mPl$ and $s \lesssim s_{\rm
max}$, where~\cite{Turner} $s_{\rm max} \simeq -[8 + 4 \log_{10}
(M/\mPl)]/[29.8 + \log_{10} (M/\mPl)]$.

One must also impose that $M \gtrsim 1\GeV$, so that the
predictions of Big Bang Nucleosynthesis are not
spoiled~\cite{Turner}.

We work in the Coulomb gauge, $A_0 = \sum_i \partial_i A_i =0$,
and we expand the electromagnetic field $A_{\mu} = (A_0, \A)$ as
\begin{equation}
\label{Aexpansion} \A(\eta,\x) = \int \!\!
\frac{d^3k}{(2\pi)^3\sqrt{2k}} \sum_{\alpha=1,2} \!
\ee_{\kk,\alpha} \, a_{\kk,\alpha} A_k(\eta) \, e^{i\kk \x} +
\mbox{h.c.},
\end{equation}
where $k = |\kk|$ and $\ee_{\kk,\alpha}$ are the transverse
polarization vectors satisfying the completeness relation
$\sum_\alpha (\varepsilon_{\kk,\alpha})_i
(\varepsilon_{\kk,\alpha}^*)_j = P_{ij}(\kk)$, with $P_{ij}(\kk) =
\delta_{ij} - k_i k_j /k^2$.
In order that the annihilation and creation operators
$a_{\kk,\alpha}$ and $a_{\kk,\alpha}^{\dag}$ satisfy the usual
commutation relations
$[a_{\kk,\alpha}, a_{\kk',\alpha'}^{\dag}] = (2\pi)^3
\delta_{\alpha \alpha'} \delta(\kk-\kk')$, we must impose the
normalization condition $A_k(\eta) \dot{A}^*_k(\eta) - A_k^*(\eta)
\dot{A}_k(\eta) = i/I$, where a dot denotes differentiation with
respect to $\eta$. This assures, in turn, that the field $A_{\mu}$
and its canonical conjugate momenta $\pi_0 = 0$ and $\pi_i = I
\dot{A}_i$ satisfy the usual commutation relations:
\begin{equation}
\label{commutation} [A_i(\eta,\x),\pi_j(\eta,\y)] = i \! \int \!\!
\frac{d^3k}{(2\pi)^3} \, e^{i\kk (\x-\y)} \, P_{ij}(\kk).
\end{equation}
The equation of motion for $A_k(\eta)$ follows from
Lagrangian~(\ref{Lagrangian}):
\begin{equation}
\label{Motion} \ddot{A}_k(\eta) + \frac{ns}{\eta} \dot{A}_k(\eta)
+ k^2 A_k(\eta) = 0.
\end{equation}
The solution of the above equation, taking into account the
normalization condition on $A_k(\eta)$, is easily found:
\begin{equation}
\label{Ainflation} A_k(\eta) = \sqrt{\frac{\pi}{2}} \,
e^{-i\pi(1+2\nu)/4} \, I^{-1/2} \, |k\eta|^{1/2} \,
H_\nu^{(2)}(|k\eta|),
\end{equation}
where $\nu = (ns-1)/2$ and $H_\nu^{(2)}(x)$ is the Hankel function
of second kind. (Observe that, since both $n$ and $s$ are
positive, it results $\nu \geq -1/2$).

We are interested in the study of large-scale electromagnetic
fields, that is in modes whose physical wavelength is much greater
than the Hubble radius, $\lambda_{\rm phys} \gg H^{-1}$ or
equivalently $|k\eta| \ll 2/|1+3\gamma|$, where $\lambda_{\rm
phys} = a \lambda$, $\lambda = 1/k$ is the comoving wavelength,
and we used $\eta = 2/[(1+3\gamma)aH]$.
Therefore, in Eq.~(\ref{Ainflation}), we can replace the Hankel
function with its small-argument expansion: $H_\nu^{(2)}(x) \simeq
i \pi^{-1} 2^\nu \Gamma(\nu) x^{-\nu}$ for $\nu > 0$,
$H_\nu^{(2)}(x) \simeq 2i \pi^{-1} \ln x$ for $\nu = 0$, and
$H_\nu^{(2)}(x) \simeq -i \pi^{-1} 2^{-\nu} \Gamma(-\nu) e^{-i\pi
\nu} x^{\nu}$ for $\nu < 0$, where $\Gamma(x)$ is the Gamma
function.

After Inflation, the Universe enters in the so-called reheating
phase, during which the energy of the inflaton is converted into
ordinary matter. In this paper, we restrict our analysis to the
case of instantaneous reheating, that is after Inflation the
Universe enters the radiation dominated era. In radiation era, the
general expression for the electromagnetic field is
\begin{equation}
\label{Aradiation} A_k^{\rm rad}(\eta) = \alpha_k e^{ik\eta} +
\beta_k e^{-ik\eta},
\end{equation}
where $\alpha_k$ and $\beta_k$ are the so-called Bogoliubov
coefficients~\cite{Birrell}, determining the spectral number
distribution of particles produced from the vacuum. By matching
expressions (\ref{Ainflation}) and (\ref{Aradiation}) and their
first derivative at the time of the end of Inflation, we find the
spectrum of the electromagnetic field generated from the vacuum at
large scales:
\begin{equation}
\label{Agenerated} |A_k^{\rm vac}(\eta_1)|^2 = |\beta_k|^2 \simeq
\frac{[2^{\nu} \Gamma(1+\nu)]^2}{2\pi I_1 \, |k\eta_1|^{1+2\nu}}
\, ,
\end{equation}
valid for $\nu \geq 0$. When $-1/2 < \nu < 0$, the electromagnetic
vacuum fluctuations go like $|A_k^{\rm vac}(\eta_1)|^2 \propto
|k\eta_1|^{1+2\nu}$ and then are vanishingly small for $|k\eta_1|
\ll 1$. In the case $\nu = -1/2$, conformal invariance is
recovered and then, as it should be, we find $|A_k^{\rm
vac}(\eta_1)|^2 = 0$, exactly, for all $|k\eta_1|$.

We can define, now, the average magnetic field on a comoving scale
$\lambda$ as~\cite{Prokopec}
\begin{equation}
\label{Blambda} B_\lambda^2(\eta) \! = a^{-4} \, \langle 0| \, |\!
\int \! d^3y \, W_\lambda(|\x-\y|) \, \nabla \! \times \!
\A(\eta,\y)|^2 \, |0 \rangle,
\end{equation}
where $W_\lambda(|\x|)$ is a suitable (real) window function, and
the vacuum state $|0\rangle$, such that $a_{\kk,\alpha} |0\rangle
= 0$, is normalized as $\langle 0|0\rangle = 1$. Taking into
account Eqs.~(\ref{Aexpansion}) and (\ref{Blambda}), we obtain
\begin{equation}
\label{Blambda1} B_\lambda^2(\eta) = \int_{0}^{\infty}\!
\frac{dk}{k} \, W_\lambda^2(k) \, \mathcal{P}_B(\eta,k),
\end{equation}
where $W_\lambda(k)$ is the Fourier Transform of the window
function and
\begin{equation}
\label{PowerB} \mathcal{P}_B(\eta,k) = \frac{k^4}{2\pi^2 a^4} \,
|A_k(\eta)|^2
\end{equation}
is the magnetic power spectrum.
As windows function we can take a gaussian window,
$W_\lambda(|\x|) = (2\pi \lambda^2)^{-3/2}
e^{-|\x|^2/(2\lambda^2)}$, so that $W_\lambda(k) = e^{-\lambda^2
k^2/2}$.

From the end of Inflation until today, due to the high
conductivity of the cosmic plasma, the magnetic field evolves
adiabatically~\cite{Turner}, $a^2 B_\lambda = \mbox{const}$, so
that $B_\lambda^{\rm today} = a_1^2 B_\lambda(\eta_1)$. Inserting
Eq.~(\ref{Agenerated}) in Eq.~(\ref{PowerB}) and taking into
account Eq.~(\ref{Blambda1}), we finally obtain:
\begin{eqnarray}
\label{B0} && \!\!\!\!\!\!\!\!\!\!\!\! B_\lambda^{\rm today} =
\frac{2^\nu \Gamma(1+\nu) [\Gamma(3/2-\nu)]^{1/2}}{(2\pi)^{3/2}
I_1^{1/2}} \, \frac{\lambda^{\nu-3/2}}{|\eta_1|^{\nu+1/2}} \, ,
\\
\label{I1} && \!\!\!\!\!\!\!\!\!\!\!\! I_1 = \left(\frac{m_{\rm
max}}{m}\right)^{\!2n} \! ,
\\
\label{eta} && \!\!\!\!\!\!\!\!\!\!\!\! |\eta_1| =
\frac{\sqrt{3}}{\sqrt{2\pi} \, |1+3\gamma|}
\left[\frac{g_{*S}(T_1)}{g_{*S}(T_0)}\right]^{\!1/3} \!
\frac{\mPl}{M T_0} \, .
\end{eqnarray}
In Eq.~(\ref{eta}), we used the fact that during radiation and
matter dominated eras the expansion parameter evolves as $a
\propto g_{*S}^{-1/3} T^{-1}$, where $T$ is the temperature and
$g_{*S}(T)$ the number of effectively massless degrees of freedom
referring to the entropy density of the Universe~\cite{Kolb}. The
temperature at the end of Inflation, the so-called reheating
temperature, is~\cite{Turner} $T_1 = M$, while $T_0$ is the actual
temperature.
\footnote{In the following we will use the values~\cite{Kolb}:
$T_0 \simeq 2.35 \times 10^{-13} \GeV$, $g_{*S}(T_0) \simeq 3.91$,
and $g_{*S}(T_1) = 106.75$ (referring to the number of effectively
massless degrees of freedom of Standard Model). It is useful to
recall that $1\G \simeq 6.9 \times 10^{-20} \GeV^2$ and $1 \Mpc
\simeq 1.56 \times 10^{38} \GeV^{-1}$.}

It is important to observe that, in order to avoid an infrared
divergence in the magnetic power spectrum, we must impose that
$\nu < 3/2$.

The ``best case scenario'', i.e. the case of maximum strength of
the magnetic field, corresponds to $s = s_{\rm max}$.

In Fig.~1, we show the actual magnetic field, in the best case
scenario, as a function of the Inflation energy scale $M$ for
different values of the index $n$ at the comoving scale $\lambda =
10\kpc$ in the two cases $m = m_e$ (upper panel) and $m = m_{\rm
max}$ (lower panel). The range of $M$ is such that $0 \leq \nu
\leq 1.49$ (both panels) and $m \lesssim m_{\rm max}$ (upper
panel).


\begin{figure}[t]
\begin{center}
\includegraphics[clip,width=0.43\textwidth]{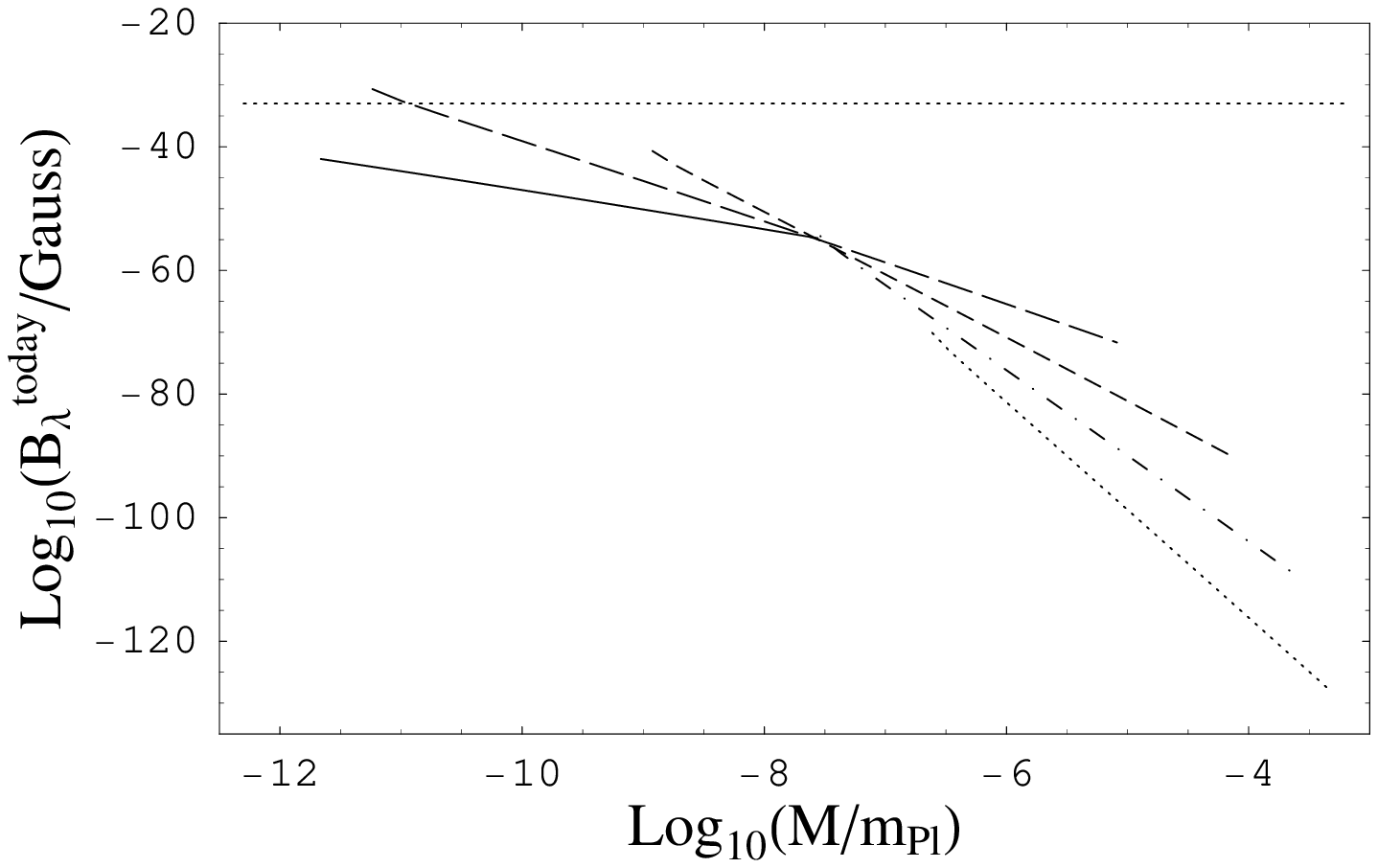}
\includegraphics[clip,width=0.43\textwidth]{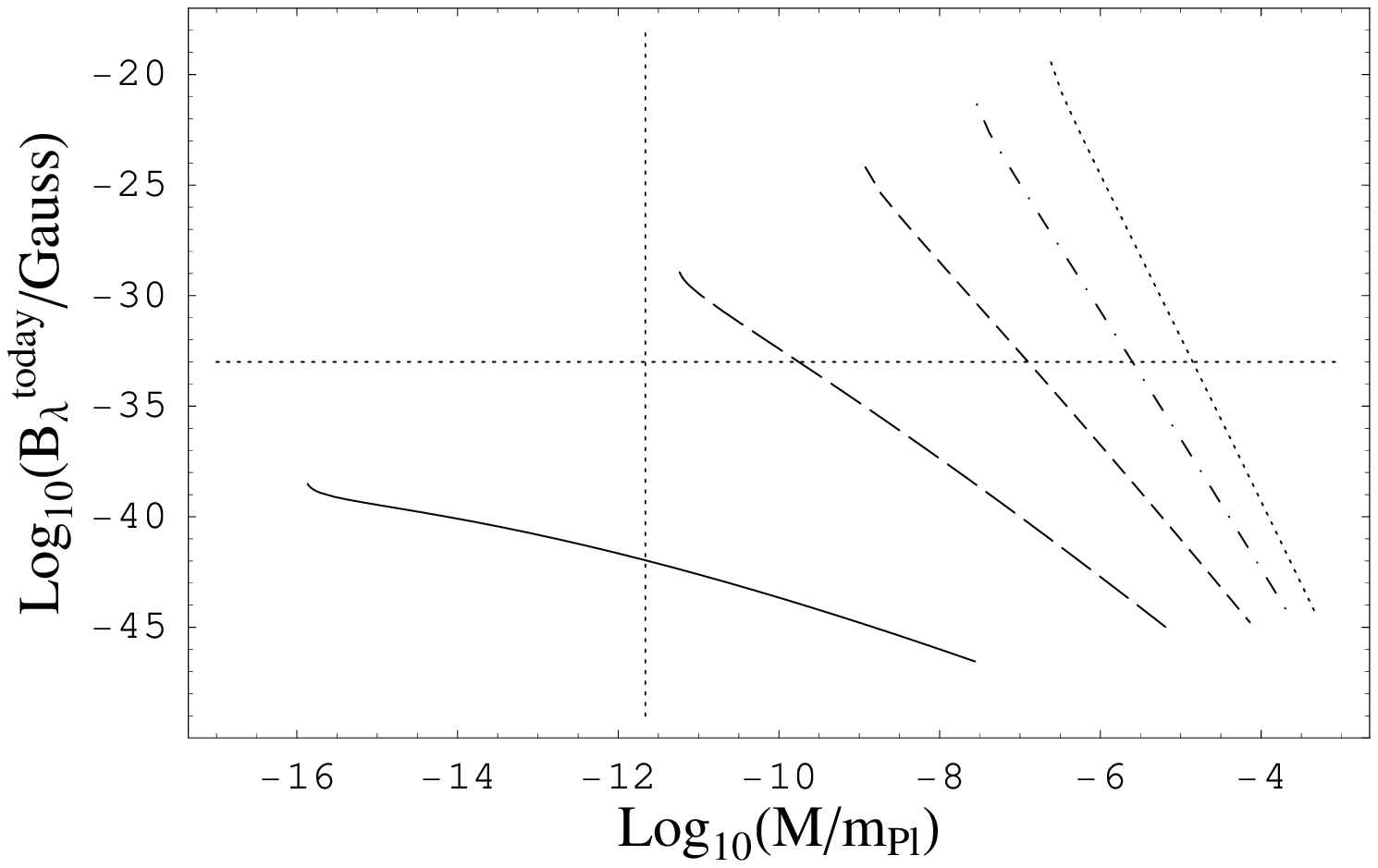}
\caption{Actual magnetic field as a function of $M$ for $m=m_e$
(upper panel) and $m=m_{\rm max}$ (lower panel) at the comoving
scale $\lambda = 10\kpc$ for different values of $n$: $n=1$
(continuous line), $n=2$ (long-dashed line), $n=3$ (dashed line),
$n=4$ (dot-dashed line), $n=5$ (dotted line). The horizontal
dotted lines refers to the minimum seed field required for dynamo
amplification, $B \simeq 10^{-33} \G$. In the lower panel, on the
left (right) side of the vertical dotted line it results $m
\lesssim m_e$ ($m \gtrsim m_e$).}
\end{center}
\end{figure}


As it is clear from the figure, if $m = m_e$ the produced fields
are not astrophysically interesting. This disagrees with the
results discussed by Mazzitelli and Spedalieri in
Ref.~\cite{Mazzitelli} where it is argued that, for adequate
values of $n$ and $\gamma$, the produced fields can serve as seed
for the cosmological magnetic field. The discrepancy resides in
the fact that, the astrophysically interesting fields they
obtained correspond, indeed, to values of $n$ and $\gamma$ (and
then of $\nu$) such that the related power spectrum is
infrared-divergent, and then are non-physical.

In the ``optimistic'' case $m = m_{\rm max}$, for all $n \geq 2$
there is a range in the parameter $M$ such that the intensity of
the produced magnetic field is greater than the minimum seed field
required for dynamo amplification, $B \gtrsim 10^{-33} \G$.


\section{III. $I F^2$ models}

We consider now a model given by Lagrangian~(\ref{Lagrangian})
with $I$ an arbitrary power function of the conformal time:
\begin{equation}
\label{I} I = I_1 \left(\frac{\eta}{\eta_1}\right)^{\!\!\alpha}
\!,
\end{equation}
where $I_1$ is a constant and we assume that $\alpha$ is a
positive real number. The above parametrization of the function
$I$, already used in the literature~\cite{Bamba}, is not the
unique admissible one. However, leaving $I$ an arbitrary function
of the conformal time, if on the one hand would render our
analysis more general, on the other hand would allow us just to
get qualitative results.

We study the production of seed fields during an inflationary
epoch described by both Power-Law and de Sitter Inflation. In the
latter case, the equation of state describing the evolution of the
Universe is $p_{\rm tot} = \gamma \rho_{\rm tot}$ with $\gamma =
-1$, so that the Hubble parameter is a constant and $\eta =
-1/(aH)$.

If $I_1 \gtrsim 1$, we can neglect the Maxwell term in
Eq.~(\ref{Lagrangian}) and then the analysis performed in Section
II applies also to the case at hand. In particular,
Eqs.~(\ref{Motion}), (\ref{Ainflation}), (\ref{B0}), and
(\ref{eta}) are still valid provided $ns$ is replaced by $\alpha$.
Consequently, now it results $\nu = (\alpha-1)/2$ and the
condition $0 \leq \nu < 3/2$ translates to $1 \leq \alpha < 4$.
The expression for the actual magnetic field can be recast as:
\begin{equation}
\label{B0deSitter} B_\lambda^{\rm today} =
\frac{10^{12\alpha-55}}{\zeta_\alpha I_1^{1/2}} \, \frac{
[\Gamma(2-\alpha/2)]^{1/2}}{|1+3\gamma|^{-\alpha/2}}
\left(\frac{M}{\mPl}\right)^{\!\!\alpha/2} \! \lambda_{\rm 10
kpc}^{(\alpha-4)/2} \G,
\end{equation}
where $\lambda_{\rm 10 kpc} = \lambda/(10 \kpc)$ and
$\zeta_\alpha$ is an increasing function of $\alpha$ of order
unity (such that $\zeta_1 \simeq 0.6$ and $\zeta_4 \simeq 2.3$).

In Fig.~2, we show the value of actual magnetic field both for
Power-Law (in the best case scenario corresponding to $s = s_{\rm
max}$) and de Sitter Inflation as a function of $M$ (which in this
case can assume all values in the interval $1 \GeV \lesssim M
\lesssim 10^{-2} \mPl$) for different values of the parameter
$\alpha$ at the comoving scale $\lambda = 10\kpc$ and for $I_1=1$.


\begin{figure}[t]
\begin{center}
\includegraphics[clip,width=0.43\textwidth]{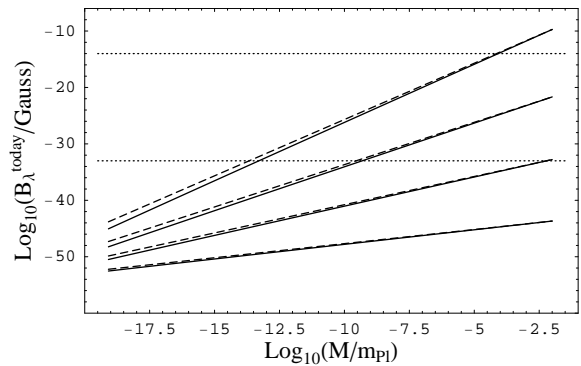}
\caption{Actual magnetic field in the case of Power-Law
(continuous lines) and de Sitter (dashed lines) Inflation as a
function of $M$ for different values of $\alpha$. From down to
top: $\alpha=1,2,3,3.99$. The first two curves from the top
correspond to $\lambda = 1\Mpc$, while the other ones to $\lambda
= 10\kpc$. The horizontal dotted lines refer to the minimum seed
fields required for dynamo amplification, $B \simeq 10^{-33} \G$,
and to directly explain galactic magnetism, $B \simeq 10^{-14}
\G$.}
\end{center}
\end{figure}


We find that for $\alpha \gtrsim 2$ there is always a minimum
value of $M$, $M_{\rm min}$, such that if $M \gtrsim M_{\rm min}$
the generated magnetic field is greater than the minimum seed
field required for a successful dynamo amplification.
Moreover, if the field is almost scale-invariant, i.e. $\alpha
\simeq 4$, and if $M$ is large enough, its intensity can be strong
enough to directly explain galactic magnetism (see the first two
curves from the top in Fig.~2 referring to $\lambda = 1\Mpc$).
In fact, expanding Eq.~(\ref{B0deSitter}) about $\alpha = 4$ and
taking $\gamma \simeq -1$ (which corresponds to take large values
of $M$), we find
\begin{equation}
\label{B0deSitter4} B_\lambda^{\rm today} \simeq \frac{2 \times
10^{-7}}{\sqrt{1-\alpha/4}} \left(\frac{M}{\mPl}\right)^{\!\!2}
\G,
\end{equation}
giving $B_\lambda^{\rm today} \gtrsim 10^{-14} \G$ for $M \gtrsim
3 (1-\alpha/4)^{1/4} 10^{15} \GeV$.


\section{III. Conclusions}

Why are large-scale magnetic fields present in all galaxies is
still a debated question in cosmology. They origin could reside in
an amplification of Inflation-produced electromagnetic vacuum
fluctuations through galactic-dynamo mechanisms.

In this paper, we have re-analyzed the production of seed magnetic
fields during Inflation considering not-conformal-invariant terms
in the electromagnetic Lagrangian of the form $(R/m^2)^n F_{\mu
\nu}F^{\mu \nu}$ and $I F_{\mu \nu}F^{\mu \nu}$.


The first model was first proposed by Turner and
Widrow~\cite{Turner} for the case $n = 1$ and then extended to the
case $n \geq 1$ by Mazzitelli and Spedalieri~\cite{Mazzitelli}.

The second model was first investigated by Bamba and
Sasaki~\cite{Bamba} in the framework of de Sitter Inflation and
assuming $I$ to be a power-law function of the conformal time
$\eta$, $I \propto \eta^\alpha$, with $\alpha$ a negative real
number.

Concerning the first model, Mazzitelli and
Spedalieri~\cite{Mazzitelli} found that, taking $m=m_e$ ($m_e$
being the electron mass), astrophysically interesting seed fields
can be generated during Inflation for adequate values of $n$. Our
analysis instead, as discussed in Section II, shows that if $m$ is
the electron mass, the produced fields are uninterestingly small
for all $n$. Nevertheless, taking $m$ as a free parameter, the
possibility to generate magnetic fields sufficiently strong in
order to seed dynamo mechanism is not excluded.

Regarding the second model, we have extended it to the case of
positive real $\alpha$ for both Power-Law and de Sitter Inflation.
We have found that, for $\alpha \gtrsim 2$, there is always a
window in the parameters defining Inflation such that the
generated magnetic fields are astrophysically interesting.
Moreover, if Inflation is (almost) de Sitter and the produced
fields almost scale-invariant ($\alpha \simeq 4$), their intensity
can be strong enough to directly explain the presence of
microgauss galactic magnetic fields.




\begin{thebibliography}{99}

\bibitem{Widrow}        For reviews on cosmic magnetic fields see:
                        L.~M.~Widrow,
                        Rev.\ Mod.\ Phys.\ {\bf 74}, 775 (2002);
                        M.~Giovannini,
                        Int.\ J.\ Mod.\ Phys.\ D {\bf 13}, 391 (2004);
                        D.~Grasso and H.~R.~Rubinstein,
                        Phys.\ Rept.\ {\bf 348}, 163 (2001).

\bibitem{LSS}           R.~E. Pudritz and J.~Silk,
                        Astrophys.\ J.\ {\bf 342}, 650 (1989);
                        A.~Lazarian,
                        Astron.\ Astrophys.\ {\bf 264}, 326 (1992);
                        K.~Subramanian, D.~Narasimha, and S.~M.~Chitre,
                        MNRAS {bf 271}, L15 (1994);
                        R.~M.~Kulsrud, {\it et al.},
                        Astrophys.\ J.\ {\bf 480}, 481 (1997);
                        G.~Davies and L.~M.~Widrow,
                        {\it ibid.} {\bf 540}, 755 (2000);
                        N.~Y.~Gnedin, A.~Ferrara, and E.~G.~Zweibel,
                        {\it ibid.} {\bf 539}, 505 (2000);
                        E.~R.~Siegel and J.~N.~Fry,
                        {\it ibid.} {\bf 651}, 627 (2006).

\bibitem{z}             P.~P.~Kronberg, J.~J.~Perry, and E.~L.~H.~Zukowski,
                        Astrophys.\ J.\ {\bf 387}, 528 (1992);
                        A.~M.~Wolfe, K.~M.~Lanzetta, and A.~L.~Oren,
                        {\it ibid.} {\bf 388}, 17 (1992);
                        R.~M.~Wolfe, {\it et al.},
                        Astron.\ Astrophys.\ {\bf 329}, 809 (1998).

\bibitem{CMB}           A.~Kosowsky and A.~Loeb,
                        Astrophys.\ J.\ {\bf 469}, 1 (1996);
                        J.~D.~Barrow, P.~G.~Ferreira, and J.~Silk,
                        Phys.\ Rev.\ Lett.\ {\bf 78}, 3610 (1997);
                        K.~Subramanian and J.~D.~Barrow,
                        {\it ibid.} {\bf 81}, 3575 (1998);
                        R.~Durrer, T.~Kahniashvili, and A.~Yates,
                        Phys.\ Rev.\ D {\bf 58}, 123004 (1998);
                        R.~Durrer, P.~G.~Ferreira, and T.~Kahniashvili,
                        {\it ibid.} {\bf 61}, 043001 (2000);
                        K.~Jedamzik, V.~Katalinic, and A.~V.~Olinto,
                        Phys.\ Rev.\ Lett.\ {\bf 85}, 700 (2000);
                        T.~R.~Seshadri and K.~Subramanian,
                        {\it ibid.} {\bf 87}, 101301 (2001);
                        A.~Mack, T.~Kahniashvili, and A.~Kosowsky,
                        Phys.\ Rev.\ D {\bf 65}, 123004 (2002);
                        L.~Campanelli, A.~D.~Dolgov, M.~Giannotti, and F.~L.~Villante,
                        Astrophys.\ J.\ {\bf 616}, 1 (2004);
                        A.~Lewis,
                        Phys.\ Rev.\ D {\bf 70}, 043011 (2004);
                        M.~Giovannini,
                        {\it ibid.} {\bf 71}, 021301 (2005);
                        A.~Kosowsky, {\it et al.},
                        {\it ibid.} {\bf 71}, 043006 (2005);
                        J.~D.~Barrow, R.~Maartens, and C.~G.~Tsagas,
                        Phys.\ Rept.\ {\bf 449}, 131 (2007), and references therein;
                        L.~Campanelli, P.~Cea and L.~Tedesco,
                        Phys.\ Rev.\ Lett.\ {\bf 97}, 131302 (2006)
                        [Erratum-ibid.\ {\bf 97}, 209903 (2006)];
                        Phys.\ Rev.\ D {\bf 76}, 063007 (2007);
                        M.~Giovannini,
                        {\it ibid.} {\bf 70}, 123507 (2004);
                        Class.\ Quant.\ Grav.\ {\bf 23}, 4991 (2006);
                        Phys.\ Rev.\ D {\bf 73}, 101302 (2006);
                        {\it ibid.} {\bf 74}, 063002 (2006);
                        PMC Phys.\ A {\bf 1}, 5 (2007)
                        [arXiv:0706.4428 [astro-ph]];
                        Phys.\ Rev.\ D {\bf 76}, 103508 (2007);
                        D.~G.~Yamazaki, K.~Ichiki, T.~Kajino, and G.~J.~Mathews,
                        Phys.\ Rev.\ D {\bf 77}, 043005 (2008).

\bibitem{CCFT}          L.~Campanelli, P.~Cea, G.~L.~Fogli, and L.~Tedesco,
                        Phys.\ Rev.\ D {\bf 77}, 043001 (2008).

\bibitem{Mazzitelli}    F.~D.~Mazzitelli and F.~M.~Spedalieri,
                        Phys.\ Rev.\ D {\bf 52}, 6694 (1995).

\bibitem{Bamba}         K.~Bamba and M.~Sasaki,
                        JCAP {\bf 0702}, 030 (2007).

\bibitem{Generation}    B.~Ratra,
                        Astrophys.\ J.\ {\bf 391}, L1 (1992);
                        W.~D.~Garretson, G.~B.~Field, and S.~M.~Carroll,
                        Phys.\ Rev.\ D {\bf 46}, 5346 (1992);
                        A.~Dolgov,
                        {\it ibid.} {\bf 48}, 2499 (1993);
                        D.~Lemoine and M.~Lemoine,
                        {\it ibid.} {\bf 52}, 1955 (1995);
                        M.~Gasperini, M.~Giovannini, and G.~Veneziano,
                        Phys.\ Rev.\ Lett.\ {\bf 75}, 3796 (1995);
                        A.~C.~Davis and K.~Dimopoulos,
                        Phys.\ Rev.\ D {\bf 55}, 7398 (1997);
                        O.~Bertolami and D.~F.~Mota,
                        Phys.\ Lett.\ B {\bf 455}, 96 (1999);
                        M.~Giovannini,
                        Phys.\ Rev.\ D {\bf 62}, 123505 (2000);
                        hep-ph/0104214;
                        Phys.\ Rev.\ D {\bf 64}, 061301 (2001);
                        T.~Prokopec,
                        astro-ph/0106247;
                        M.~Gasperini,
                        Phys.\ Rev.\ D {\bf 63}, 047301 (2001);
                        B.~A.~Bassett, G.~Pollifrone, S.~Tsujikawa and F.~Viniegra,
                        {\it ibid.} {\bf 63}, 103515 (2001);
                        K.~Dimopoulos, T.~Prokopec, O.~Tornkvist, and A.~C.~Davis,
                        {\it ibid.} {\bf 65}, 063505 (2002);
                        K.~Bamba and J.~Yokoyama,
                        Phys.\ Rev.\ D {\bf 69}, 043507 (2004);
                        {\it ibid.} {\bf 70}, 083508 (2004);
                        A.~Ashoorioon and R.~B.~Mann,
                        {\it ibid.} {\bf 71}, 103509 (2005);
                        C.~G.~Tsagas,
                        {\it ibid.} {\bf 72}, 123509 (2005);
                        K.~E.~Kunze,
                        Phys.\ Lett.\ B {\bf 623}, 1 (2005);
                        M.~R.~Garousi, M.~Sami and S.~Tsujikawa,
                        {\it ibid.} {\bf 606}, 1 (2005);
                        M.~M.~Anber and L.~Sorbo,
                        JCAP {\bf 0610}, 018 (2006);
                        A.~Akhtari-Zavareh, A.~Hojati, and B.~Mirza,
                        Prog.\ Theor.\ Phys.\ {\bf 117}, 803 (2007);
                        K.~E.~Kunze,
                        arXiv:0710.2435 [astro-ph];
                        M.~Giovannini,
                        Phys.\ Lett.\ B {\bf 659}, 661 (2008);
                        J.~Martin and J.~Yokoyama,
                        JCAP {\bf 0801}, 025 (2008).

\bibitem{Prokopec}      T.~Prokopec and E.~Puchwein,
                        Phys.\ Rev.\ D {\bf 70}, 043004 (2004).

\bibitem{Turner}        M.~S.~Turner and L.~M.~Widrow,
                        Phys.\ Rev.\ D {\bf 37}, 2743 (1988).

\bibitem{Dynamo}        For recent reviews on dynamo mechanisms see, e.g.:
                        A.~Brandenburg and K.~Subramanian,
                        Phys.\ Rept.\ {\bf 417}, 1 (2005);
                        A.~Shukurov,
                        astro-ph/0411739.

\bibitem{Birrell}       N.~D.~Birrell and P.~C.~W.~Davies,
                        {\it Quantum Fields in Curved Space}
                        (Cambridge University Press, New York, 1982).

\bibitem{Kolb}          E.~W.~Kolb and M.~S.~Turner,
                        {\it The Early Universe}
                        (Addison-Wesley, Redwood City, California, 1990).

\end{thebibliography}
\end{document}